\documentclass[a4paper]{jpconf}
\usepackage{graphicx}
\begin{document}
\title{$A$-site substitution effect on physical properties of Sr$_3$Fe$_{2-x}$Co$_x$O$_{7-\delta}$}

\author{J. Tozawa, M. Akaki, D. Akahoshi, H. Kuwahara, and K. Itatani}

\address{Faculty of Science and Technology, Sophia University, Tokyo 102-8554, Japan}

\ead{j-tozawa@sophia.ac.jp}

\begin{abstract}
We have investigated the $Ln^{3+}$-substitution ($Ln$ = lanthanoid) effect of a quasi two-dimensional ferromagnet Sr$_3$Fe$_{2-x}$Co$_x$O$_{7-\delta}$ ($x$$ = $$0.5$).
With increasing $Ln^{3+}$-concentration, the ferromagnetism is gradually suppressed and the resistivity is increasing, which are ascribed to an increase in antiferromagnetic (AFM) clusters created by $Ln^{3+}$-substitution.
In Sr$_{2.7}$Gd$_{0.3}$Fe$_{1.5}$Co$_{0.5}$O$_{7-\delta}$, the magnetoresistance (MR) is enhanced by about 20 \% compared with that of Sr$_3$Fe$_{1.5}$Co$_{0.5}$O$_{7-\delta}$.
Coexistence of ferromagnetic (FM) and AFM phases is essential for the enhancement of the MR\@.
Applied magnetic fields align the FM clusters in the same direction, resulting in a reduction in the resistivity.
A metamagnetic transition observed in the $Ln^{3+}$-doped samples also contributes to the enhancement of the MR.

\end{abstract}

\section{Introduction}
Since the discovery of the colossal magnetoresistance (CMR) effect, Mn oxides with a perovskite structure and their derivatives have been intensively investigated.
In particular, two-dimensional ferromagnetic (FM) metal (La,Sr)$_3$Mn$_2$O$_7$ has been attracting much attention because such metals are rarely reported.
Sr$_3$Fe$_{2-x}$Co$_x$O$_{7-\delta}$ can be expected as one of such rare materials[1,~2].
The end material Sr$_3$Fe$_2$O$_{7-\delta}$ is an antiferromagnet below $T_{\rm N}$ = 120 K, and changes from a semiconductor to a metal at 350 K[3].
With increasing Co-concentration, this system approaches an FM metal and negative magnetoresistance (MR) is observed.
Ghosh $et$ $al$.\ suggest that the ferromagnetism is driven by spin-dependent electron transfer processes between Fe$^{4+}$ and Co$^{4+}$[4].
Since highly oxidized states of both Fe$^{4+}$ and Co$^{4+}$ are unstable, the crystal structure of Sr$_3$Fe$_{2-x}$Co$_x$O$_{7-\delta}$ has a large amount of oxygen vacancies.
The oxygen deficiency $\delta$ of Sr$_3$Fe$_{1.4}$Co$_{0.6}$O$_{7-\delta}$ treated under high oxygen pressure of 600 bar at 1173 K is 0.20 $\pm$ 0.01[4].
A decrease in the average valence of $B$-site cations weakens the spin-dependent electron transfer process, which probably leads to coexistence of FM and AFM (or PM) phases.
Such phase coexistence can be expected to enhance the MR of Sr$_3$Fe$_{2-x}$Co$_x$O$_{7-\delta}$, as is often the case with CMR manganites. 
In~this study, in order to enhance the MR, we have prepared $Ln$-substituted Sr$_3$Fe$_{1.5}$Co$_{0.5}$O$_{7-\delta}$ and have investigated the physical properties.

\section{Experiment}
Sr$_{3-y}Ln_y$Fe$_{1.5}$Co$_{0.5}$O$_{7-\delta}$ was prepared in a polycrystalline form by a solid-state reaction method.
Mixed powders of SrCO$_3$, $Ln_2$O$_3$ ($Ln$ = La, Nd, and Gd), Fe$_3$O$_4$, and CoO were heated at 1173 K in air and sintered at 1673 K in O$_2$. 
The sample was then annealed under high oxygen pressure of 240 bar at 773~K for 10 h, and cooled down to about 300 K at a rate of 40 K/h.
Crystallographic analysis of the obtained samples was performed by X-ray diffraction method at room temperature.
The oxygen content of the synthesized samples was estimated using iodometric titration with an accuracy of $\pm$ 0.02.
The magnetic properties were measured using a Quantum Design, Physical Property Measurement System (PPMS - 9T).
The resistivity was measured using a standard four-probe method.

\section{Results and discussion} 

\begin{figure}
\begin{center}
\includegraphics[width=0.98 \textwidth,clip]{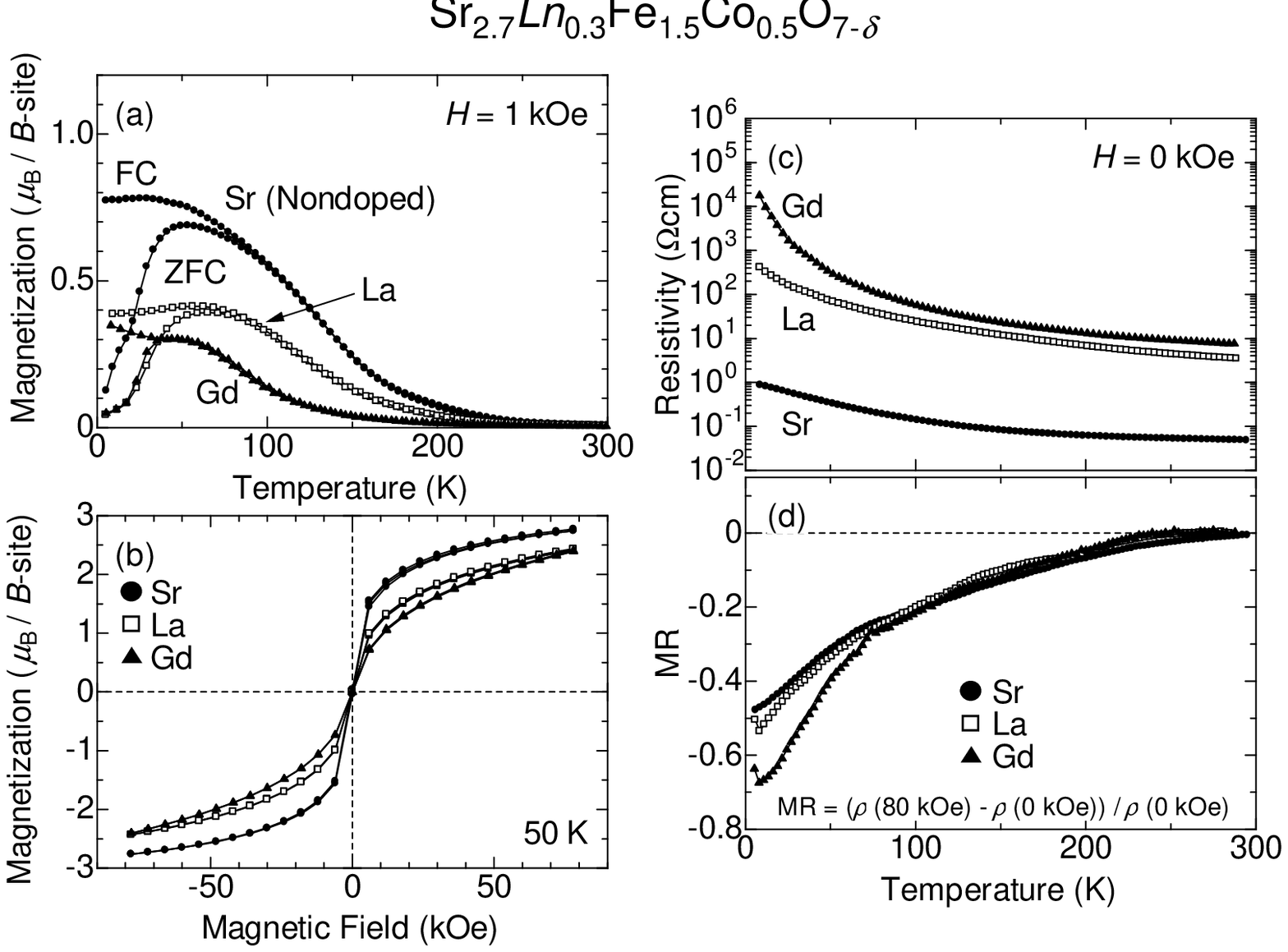}
\end{center}
\caption{(a) Temperature and (b) magnetic field dependence of magnetization of Sr$_{2.7}Ln_{0.3}$Fe$_{1.5}$Co$_{0.5}$O$_{7-\delta}$ ($Ln$ = La and Gd). A nondoped sample is also presented for comparison. Temperature dependence of (c) resistivity in a zero field and (d) magnetoresistance (MR). Here the MR is defined as MR = ($\rho$(80 kOe) $-$ $\rho$(0 kOe)) / $\rho$(0 kOe).
}
\end{figure}

Figures 1 (a) and (b) show the temperature dependence of the magnetization ($M$-$T$) and magnetization curves ($M$-$H$) of Sr$_{2.7}Ln_{0.3}$Fe$_{1.5}$Co$_{0.5}$O$_{7-\delta}$ ($Ln$ = La and Gd).
The $M$-$T$ of Sr$_3$Fe$_{1.5}$Co$_{0.5}$O$_{7-\delta}$ shows an increase below 250 K\@.
The zero field cooling (ZFC) process deviates from field cooling (FC) below 80 K\@.
The magnetization is FM below 250 K, and the magnetization value reaches about 2.8 $\mu_{\rm B}$/$B$-site at 50 K, as seen from the $M$-$H$ curve.
These magnetic behaviors result from the emergence of the cluster glass state, as previously reported[1].
This cluster glass-like behavior is prominent below 80 K\@.
In~Sr$_{2.7}$La$_{0.3}$Fe$_{1.5}$Co$_{0.5}$O$_{7-\delta}$, the onset temperature of the FM transition shifts to lower temperatures, and the FM moment is slightly suppressed (Fig.\ 1(b)).
In the case of Gd$^{3+}$-substitution, the FM moment is close to that of Sr$_{2.7}$La$_{0.3}$Fe$_{1.5}$Co$_{0.5}$O$_{7-\delta}$, but the FM transition temperature is further lower to around 150 K.

Figures 1 (c) and (d) show the temperature dependence of the resistivity and MR of Sr$_{2.7}Ln_{0.3}$Fe$_{1.5}$Co$_{0.5}$O$_{7-\delta}$.
Here the MR value is defined as MR = ($\rho$(80 kOe) $-$ $\rho$(0 kOe)) / $\rho$(0~kOe).
The resistivity of Sr$_3$Fe$_{1.5}$Co$_{0.5}$O$_{7-\delta}$ is semiconducting, and the negative MR is observed below the onset temperature of ferromagnetism.
The absolute value of MR,~$\bigl|$MR$\bigr|$, increases gradually up to $50$ \% at 5~K with decreasing temperature.
La$^{3+}$-substitution makes this system insulating, while the MR of $Ln$ = La is slightly enhanced compared with that of Sr$_3$Fe$_{1.5}$Co$_{0.5}$O$_{7-\delta}$ at low temperatures.
With decreasing the ionic radius of $Ln^{3+}$, the resistivity is increasing and the $\bigl|$MR$\bigr|$ is further enhanced from $47$ \% (nondoped) to $68$ \% ($Ln$~=~Gd).

\begin{figure}[tb]
\begin{minipage}{0.50 \textwidth}
\begin{center}
\includegraphics[width=0.98 \textwidth,clip]{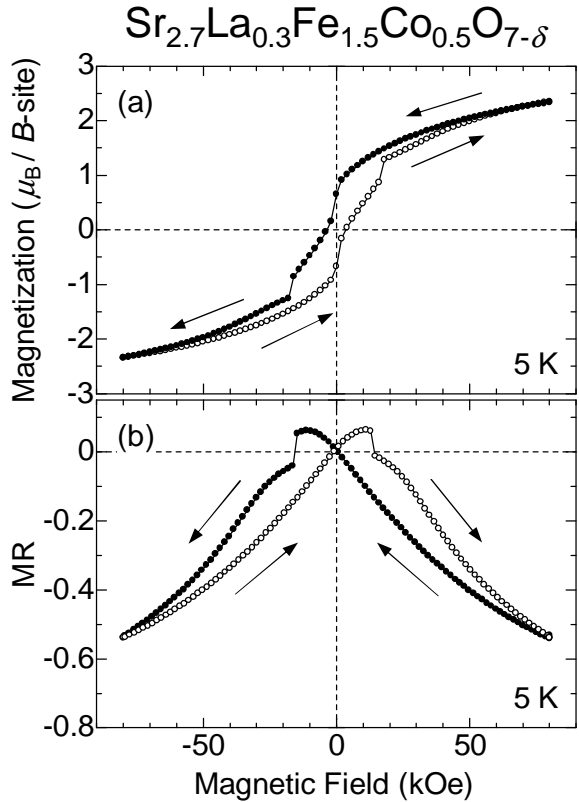}
\end{center}
\end{minipage}
\begin{minipage}{0.44 \textwidth}
\begin{center}
\caption{Magnetic field dependence of (a) magnetization and (b) magnetoresistance (MR) of Sr$_{2.7}$La$_{0.3}$Fe$_{1.5}$Co$_{0.5}$O$_{7-\delta}$ at 5 K\@. The arrows mean the scan directions.}
\end{center}
\end{minipage}
\end{figure}

Figures 2 (a) and (b) show the $M$-$H$ curve and isothermal MR of Sr$_{2.7}$La$_{0.3}$Fe$_{1.5}$Co$_{0.5}$O$_{7-\delta}$ at 5~K\@.
A significant difference between La$^{3+}$-doped and nondoped samples is that a metamagnetic transition is found in the La$^{3+}$-doped sample around 16~kOe.
The metamagnetic transition accompanies an abrupt decrease in the resistivity and is reversibly observed by applying magnetic fields.
Similar metamagnetic behavior is observed in all the $Ln^{3+}$-substituted samples, but not in the nondoped sample.

We discuss the origin of the enhancement of the MR of Sr$_{2.7}Ln_{0.3}$Fe$_{1.5}$Co$_{0.5}$O$_{7-\delta}$.
We confirmed that the $\delta$ of the $Ln^{3+}$-substituted sample almost coincides with that of Sr$_3$Fe$_{1.5}$Co$_{0.5}$O$_{7-\delta}$ by iodometric titration.
In other words, $Ln^{3+}$-substitution merely decreases the average valence of the $B$-site cation.
An oxygen-vacancy ordered structure (or oxygen stoichiometry) might exist in the vicinity of $\delta$ = 0.20.
Such a decrease in the average valence of the $B$-site cation probably creates antiferromagnetic (AFM) (or non-FM) clusters around $Ln^{3+}$.
It is therefore reasonable to conclude that the suppression of ferromagnetism by $Ln^{3+}$-substitution is due to the creation of these AFM (or non-FM) clusters.
The volume reduction of the FM clusters and randomness due to $Ln^{3+}$-substitution cause the resistivity to be insulating.
The FM clusters are aligned parallel to applied magnetic fields.
As a result, the large negative MR is observed.
In addition, metamagnetism accompanying a~reduction in the resistivity also contributes to the enhancement of the MR at low temperatures.

\section{Summary}
We have investigated a quasi two-dimensional ferromagnet Sr$_3$Fe$_{1.5}$Co$_{0.5}$O$_{7-\delta}$ partially substituted by $Ln^{3+}$\@.
This substitution of $Ln^{3+}$ suppresses ferromagnetism and causes the system to be more insulating, which are attributed to an increase in AFM clusters created around $Ln^{3+}$\@.
In Sr$_{2.7}$Gd$_{0.3}$Fe$_{1.5}$Co$_{0.5}$O$_{7-\delta}$, the MR value is enhanced by about 20 \% compared with that of Sr$_3$Fe$_{1.5}$Co$_{0.5}$O$_{7-\delta}$.
Competition between FM and AFM (or non-FM) clusters is essential for the enhancement of the MR\@.
One of the plausible explanations for the origin of this enhanced MR is as follows.
When applying magnetic fields to Sr$_{2.7}$Gd$_{0.3}$Fe$_{1.5}$Co$_{0.5}$O$_{7-\delta}$, the FM clusters are aligned in the same direction, resulting in a reduction in resistivity.
In addition, in the $Ln^{3+}$-doped samples, a metamagnetic transition accompanying an abrupt decrease in the resistivity occurs at low temperatures, which also contributes to the enhancement of the MR of Sr$_{2.7}Ln_{0.3}$Fe$_{1.5}$Co$_{0.5}$O$_{7-\delta}$.

\section*{Acknowledgments}
This work was partly supported by the Asahi Glass Foundation, and Grant-in-Aid for scientific research (C) from the Japan Society for Promotion of Science.

\end{document}